\documentclass{amsart}
\usepackage{amsfonts}

\usepackage{amsmath,amsfonts,amssymb}
\usepackage[notref,notcite]{showkeys}
\usepackage{graphicx}

\input tcilatex
\theoremstyle{plain}

\theoremstyle{definition}

\numberwithin{equation}{section}
\def\beq{\begin{equation}}
\def\eeq{\end{equation}}
\def\beqa{\begin{eqnarray}}
\def\eeqa{\end{eqnarray}}
\def\12{\textstyle{1\over2}}

\input tcilatex

\begin{document}
\title[Metastable wetting]{Metastable wetting}
\author{Jo\"{e}l de Coninck $^{\left( 1\right) }$, Francois Dunlop $^{\left(
2\right) }$, Thierry Huillet $^{\left( 2\right) }$}
\address{$^{\left( 1\right) \text{ }}$Laboratoire de Physique des Surfaces et des
Interfaces\\
Universit\'{e} de Mons-Hainaut, 20 Place du Parc, 7000 Mons, Belgium\\
$^{\left( 2\right) }$ Laboratoire de Physique Th\'{e}orique et
Mod\'{e}lisation\\
CNRS-UMR 8089 et Universit\'{e} de Cergy-Pontoise\\
2 Avenue Adolphe Chauvin, 95032, Cergy-Pontoise, France\\
E-mail:\\
joel.deconinck@umons.ac.be\\
Francois.Dunlop@u-cergy.fr\\
Thierry.Huillet@u-cergy.fr}
\maketitle

\begin{abstract}
Consider a droplet of liquid on top of a grooved substrate. The wetting or
not of a groove implies the crossing of a potential barrier as the interface
has to distort, to hit the bottom of the groove. We start with computing the
free energies of the dry and wet states in the context of a simple
thermodynamical model before switching to a random microscopic version
pertaining to the Solid-on-Solid (SOS) model. For some range in parameter
space (Young angle, pressure difference, aspect ratio), the dry and wet
states both share the same free energy, which means coexistence. We compute
these coexistence lines together with the metastable regions. In the SOS
case, we describe\emph{\ }the dynamic transition between coexisting states
in wetting. We show that the expected time to switch from one state to the
other grows exponentially with the free energy barrier between the stable
states and the saddle state, proportional to the groove's width. This random
time appears to have an exponential-like distribution.\newline


\end{abstract}

\section{Introduction and outline}

The detailed study of superhydrophobic surfaces has revealed that on a rough
substrate, a drop can present two shapes: either one obeying the
Cassie-Baxter equation or one obeying the Wenzel equation \cite{Bo}, \cite
{Gross}, \cite{Bor}. On the other hand, any transition between two states
depends on the height of the barrier which has to be overcome. The
corresponding transition will thus be a function of time as revealed already
by Kramers law. We review here this crucial aspect of the problem within the
framework of an exactly solvable statistical mechanics model.\emph{\ }

\begin{figure}[tbp]
\centerline{\ \mbox{\includegraphics[width=8cm]{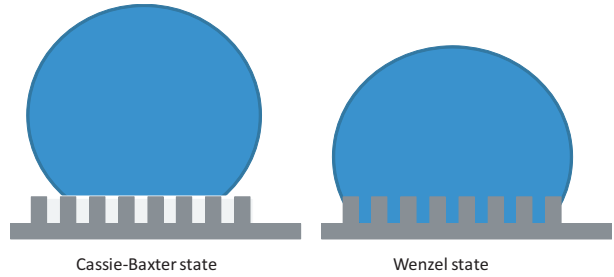}}}
\caption{The Cassie-Wenzel states}
\label{cassiewenz}
\end{figure}

\begin{figure}[tbp]
\centerline{\ \mbox{\includegraphics[width=8cm]{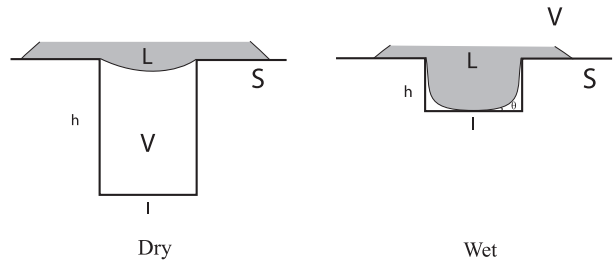}}}
\caption{Dry versus Wet states}
\label{dry_wet}
\end{figure}

The corresponding Cassie-Baxter and Wenzel states are illustrated in Fig. 
\ref{cassiewenz}. In the first (dry) state, there is no wetting of the sides
of the U-shaped wells, with vapor trapped in-between, whereas in the second
(wet) one, there is at least a partial wetting of the bottom of the U, with
vapor trapped in the corners. This is illustrated in Fig. \ref{dry_wet},
zooming on a single well. The transition between these two states is related
to the amplitude of some free-energy barrier: in this rough picture, this
barrier can be overcome by applying an external force triggered for instance
by an impact velocity or equivalently by increasing the pressure difference $%
\Delta p$ between liquid and vapor. A classical way to characterize the wet
shape is given by the Young contact angle $\theta ,$ as sketched on Fig. \ref
{dry_wet}. \newline

A rough substrate is thus made of clefts (wells, grooves...) governing
wetting properties. For a better understanding of this physical problem, we
need first to analyze and characterize the free energies of both states as
functions of the parameters $\theta $, $\Delta p$ and $\rho $, the latter
being the shape factor of a typical well. We first do that in Section $2$
for two simple macroscopic solvable models of wetting, one simpler without
the pressure parameter and the other one including pressure. For the
isotropic model including pressure, we show that there exists a region in
parameter space where the two states can both exist (the free energy
function has two stable minima) and therefore regions where only a single of
these two states exists (the free energy function possesses a single stable
minimum). Within the metastable region of existence of the two states, we
exhibit a line of coexistence along which the free energies of both states
are exactly equal.\newline

We can easily imagine that, due to fluctuations, there will be an
opportunity to flip between the `dry' and `wet' states of Fig. \ref{dry_wet}%
. So we need to embed our wetting problem into the framework of a stochastic
model of wetting. For this purpose, we used the microscopic statistical
mechanics SOS model. In this context also, the wetting of a well implies the
crossing of a potential barrier: the interface has to distort, to hit the
bottom of the well. We wish to understand this problem in more details. 
\newline

The aim of Section $3$ is to show that we are still able to compute
analytically the equilibrium free energies of the dry and wet phases in this
statistical SOS context. Let us be more precise.\emph{\ }

We study our wetting problem in the context of the SOS model in a square
well of width $l=:nl_{0}$ and height $h,$ with $h/l=:\rho $. The equilibrium
measure for the interface heights $h_{i}\geq 0$ is given by: 
\begin{equation}
d\mu _{n}=Z_{n}^{-1}e^{-\left( J\sum_{i=0}^{n}\left| h_{i+1}-h_{i}\right| +%
\frac{K}{n}\sum_{i=1}^{n}h_{i}\right) }\prod_{i=0}^{n}\left( 1+a\delta
\left( h_{i}\right) \right) dh_{i},  \label{1}
\end{equation}
with $h_{0}=h_{n+1}=h$. Here $Z_{n}$ is the normalizing partition function,
and the pressure term comes from $\Delta p\int_{0}^{l}h\left( x\right)
dx\simeq \Delta p\sum_{i}h_{i}\frac{l}{n}=\frac{K}{n}\sum h_{i}$, which
defines $K$ as $K:=l\Delta p.$ The length $l_{0}=l/n$\ is the coarse grain
length at which the SOS model is defined as an approximation to a truly
microscopic model. Thanks to this coarse-graining, the interface, originally
of finite-width, constrained by the corners of the well, has become a
surface pinned at the corners.\newline

The SOS model has three positive parameters $\left( J,K,a\right) $,
representative for the first two of the length of the interface and the area
below the interface, and for the third, of the \emph{liquid's} affinity for
the bottom of the well. More precisely, 
\begin{equation*}
-\frac{\partial \log Z_{n}}{\partial J}=\left\langle \sum_{i=0}^{n}\left|
h_{i+1}-h_{i}\right| \right\rangle ,\text{ }-l_{0}\frac{\partial \log Z_{n}}{%
\partial \left( K/n\right) }=l_{0}\left\langle
\sum_{i=1}^{n}h_{i}\right\rangle ,\text{ }l_{0}\frac{\partial \log Z_{n}}{%
\partial a}=l_{0}\left\langle \sum_{i=0}^{n}\delta \left( h_{i}\right)
\right\rangle 
\end{equation*}
are respectively the mean excess length of interface, the mean area below
the interface, and the mean wetted length of well. These three parameters
are in correspondence with the surface tension $\sigma _{L,V}$ between the liquid 
and vapor phases, the pressure difference $\Delta p$\ and the Young angle $\theta .$

In the thermodynamic limit, the structure of the interfaces is given by a
Wulff shape pinned at the corners of the well, expressed analytically in
terms of the projected surface tension. Two cases arise (see Fig. \ref
{dry_wet}):\newline

- `dry' case: the Wulff shape does not hit the bottom of the well and so
hangs between the two corners.

- `wet' case: the Wulff shape does hit the bottom of the well and the
interface is made of three pieces whose central part, flat (corresponding to
the wetting of the substrate by the fluid), is linked up to the corners of
the well by two symmetric pieces of the Wulff type. \newline

For the sake of simplicity, in this study, the vertical sides of the well
are completely hydrophobic. This fits with the SOS model (\ref{1}) with $%
h_{i}\in \Bbb{R}$\ for which the Wulff shape has no vertical part.

Using the results obtained in \cite{AS1}-\cite{BN}, \cite{CD}-\cite{CDH}, we
can compute exactly the normalized free energy for each of these two
configurations. At fixed $J$ and $\rho $, we compute the line of coexistence 
$a=a\left( K\right) $ for which these free energies coincide. Due to a
one-to-one correspondence between $a$ and $\theta $, a phase diagram $\theta
=\theta \left( K\right) $ follows. In this phase diagram, we compute the
lines separating a metastable region where the two phases coexist and stable
regions where a single phase (either wet or dry) is stable.\newline

When both `wet' and `dry' SOS states share the same free energies, we are
subsequently interested by the transition between them. In the Monte-Carlo
dynamics of Section $4$, we show that the system undergoes rare transitions
between these two equilibrium states as time passes by. We also study the
first passage times from one state to the other, together with finite-size
corrections. This requires some preliminary understanding of the free energy
of an unstable SOS saddle state which can be computed explicitly, in the
thermodynamic limit. We show that the expected time to switch from one state
to the other grows exponentially with the free energy barrier between the
stable states and the saddle state, proportional to the system's size. This
random time turns out to have an exponential-like distribution.

In Appendix C, a toy Markov-chain model is designed which illustrates the
typical behaviors encountered in our random wetting problem. The rate of
growth of the expected time to move from one stable state to the other when
an unstable state lies in-between is the energy barrier between the unstable
and stable states (Kramers' law), and the random time itself normalized by
its mean converges in distribution to an exponential probability
distribution.

\section{A simple macroscopic model. Equilibrium free energies}

Before we run into similar considerations pertaining to the SOS model, let
us first briefly describe some easy macroscopic arguments that led us to the
forthcoming study.

\subsection{No pressure $(K=0)$\newline
}

DRY PHASE. Consider a well of length $l$ and height $h=\rho l.$ Call it the
substrate ($S$). Fix the origin of the $\left( x,z\right) $ axis at the
middle of the bottom of the well. In the following, we take $l$\ as unit of
length, so that the bottom of the well is the interval $\left[
-0.5,+0.5\right] $\ at height $z=0.$

The well is initially filled with gas or vapor ($V$). We wish to fill this
groove with a liquid ($L$) in the absence of pressure ($\Delta p=\frac{K}{l}%
=0$). We assume that the vertical walls of the well remain dry. In a dry
phase, there exists a Wulff shape separating the $LV$ phases which is pinned
at $\left( x=-0.5,z=\rho \right) $ and $\left( x=0.5,z=\rho \right) $
without hitting the bottom of the well: it is the straight line joining
these two corner points. The well is entirely filled with vapor, the liquid
entirely stands above this straight separating line and so the liquid does
not wet the substrate at all. The specific free energy of this dry phase is 
\begin{equation}
f_{dry}:=\frac{F_{dry}}{l}=\sigma _{LV}+\left( 1+2\rho \right) \cdot \sigma
_{VS},  \label{I1}
\end{equation}
where $\sigma _{AB}$ is the surface tension between the phases $A$ and $B.$ 
\newline

WET PHASE $(K=0)$. In a wet phase, the liquid will meet the substrate at the
bottom of the well. Let $\theta $ be the Young angle between the vapor phase
and the substrate at the right-most meeting point. The larger $\theta $ is,
the more the substrate is hydrophilic.

Let $x_{1}\in \left( 0,0.5\right) $. In a wet phase with $K=0$, the Wulff
shape would indeed be made of two symmetric linear pieces of length $\frac{%
\rho }{\sin \theta }$, separating phase $L$ from $V$, one joining point $%
\left( x=-0.5,z=\rho \right) $ to point $\left( x=-x_{1},z=0\right) $ and
the other joining $\left( x=x_{1},z=0\right) $ to $\left( x=0.5,z=\rho
\right) .$ In between, the line joining $\left( x=-x_{1},z=0\right) $ to $%
\left( x=x_{1},z=0\right) $ would be the flat wetting zone for the
substrate. For a given $\theta $, a wet phase can exist if and only if $%
x_{1}>0$ or else 
\begin{equation}
\rho <0.5\tan \theta  \label{I2}
\end{equation}
(the well is not too deep).

When this wet phase exists, its specific free energy is 
\begin{equation}
f_{wet}:=\frac{F_{wet}}{l}=2\frac{\rho }{\sin \theta }\cdot \sigma
_{LV}+\left( 1+2\rho -2x_{1}\right) \cdot \sigma _{VS}+2x_{1}\cdot \sigma
_{LS}.  \label{I3}
\end{equation}
When both phases exist, the question of which phases is favored makes sense.
So we can ask for conditions under which $F_{dry}\leq F_{wet}$. Using $%
\sigma _{LS}-\sigma _{VS}=\sigma _{LV}\cos \theta $, we easily get the
condition 
\begin{equation}
\rho \geq 0.5\tan \left( \theta /2\right) ,  \label{I4}
\end{equation}
meaning that for the dry phase to win over the wet one, the depth of the
well has to be large enough.

So, when a wet phase exists ($\rho <0.5\tan \theta $), the dry phase wins
over the wet phase whenever the well's depth satisfies $\rho \geq 0.5\tan
\left( \theta /2\right) $. Else, if $\rho >0.5\tan \theta $, this problem
does not make sense simply because only the dry phase exists and so it
necessarily wins. So, in this oversimplified wetting problem, we expect a
parameter range for which the two phases wet and dry can coexist (a
metastable region), and within this parameter range another parameter range
for which one phase is favorable over the other. When $\rho =0.5\tan \left(
\theta /2\right) $, the two phases share the same free energy (a bistable
line of coexistence of both phases).

\subsection{Pressure $(K>0)$}

WET PHASE. The Wulff-shaped lines separating the $LV$ phases are now
symmetric convex circle arcs with radius 
\begin{equation*}
r=\frac{\sigma _{LV}}{\Delta p}=l\frac{\sigma _{LV}}{K}.
\end{equation*}
Define the scaled radius as $R=r/l.$ In the scaled length unit and in the
wet phase, we have one circle arc joining point $\left( x=-0.5,z=\rho
\right) $ to point $\left( x=-x_{1},z=0\right) $ and the other joining $%
A:=\left( x=x_{1},z=0\right) $ to $B:=\left( x=0.5,z=\rho \right) .$

Let $O$ be the center of the latter circle. Let $\varphi $ be the angle $%
\left( OA,OB\right) .$ Then the scaled euclidean distance between $A$ and $B$
is given by (see Appendix A1) 
\begin{equation*}
AB=\left[ 2R^{2}\left( \sin ^{2}\theta +\frac{\rho \cos \theta }{R}-\sqrt{%
\Delta }\right) \right] ^{1/2}.
\end{equation*}
where $\Delta =\left( \sin ^{2}\theta +\frac{\rho \cos \theta }{R}\right)
^{2}-\left( \frac{\rho }{R}\right) ^{2}$. Next $\widehat{AB}=2R\arcsin
\left( AB/\left( 2R\right) \right) $ is the scaled arc length of the arc
joining $A$ to $B.$ Using 
\begin{equation*}
AB=\sqrt{\left( 0.5-x_{1}\right) ^{2}+\rho ^{2}}.
\end{equation*}
gives $x_{1}$ as an explicit function of $\left( \theta ,\rho ,R\right) $.%
\newline

For a given $\theta $, the wet phase can exist if and only if $x_{1}>0$
which is, observing $\tan \left( \theta +\varphi /2\right) =\frac{\rho }{%
0.5-x_{1}}$: 
\begin{equation*}
\rho <0.5\tan \left( \theta +\varphi /2\right)
\end{equation*}
(again, the well should not be too deep).\newline

When this wet phase exists, its specific free energy is found to be 
\begin{equation}
f_{wet}:=\frac{F_{wet}}{l}=2\widehat{AB}\cdot \sigma _{LV}+\left( 1+2\rho
-2x_{1}\right) \cdot \sigma _{VS}+2x_{1}\cdot \sigma _{LS}+2K\cdot A_{wet},
\label{I5}
\end{equation}
where $A_{wet}=\frac{\rho }{2}\left( 0.5-x_{1}\right) -\frac{1}{2}\left[
R^{2}\varphi -R^{2}\sin \varphi \right] $ is the scaled dry area of the
vapor beneath the right part of the Wulff shape$.$ The right-most pressure
term in (\ref{I5}) comes from 
\begin{equation*}
\frac{1}{l}\left( \Delta p\cdot l^{2}A_{wet}\right) =K\cdot A_{wet}.
\end{equation*}

DRY PHASE. With $\widehat{B^{\prime }B}=2R\arcsin \left( \frac{1}{2R}\right)
=:R\psi ,$ the arc length between the left and right corner points $%
B^{\prime }$ and $B$ of the well, $f_{wet}$ should be compared to the
specific free energy in the dry phase which is 
\begin{equation}
f_{dry}:=\frac{F_{dry}}{l}=\widehat{B^{\prime }B}\cdot \sigma _{LV}+\left(
1+2\rho \right) \cdot \sigma _{VS}+KA_{dry}.  \label{I6}
\end{equation}
Here $A_{dry}=\rho -\frac{1}{2}\left[ R^{2}\psi -R^{2}\sin \psi \right] $ is
the scaled dry area of the vapor beneath the hanging Wulff line anchored at $%
B^{\prime }$ and $B.$\newline

\begin{figure}[tbp]
\centerline{\ \mbox{\includegraphics[width=12cm]{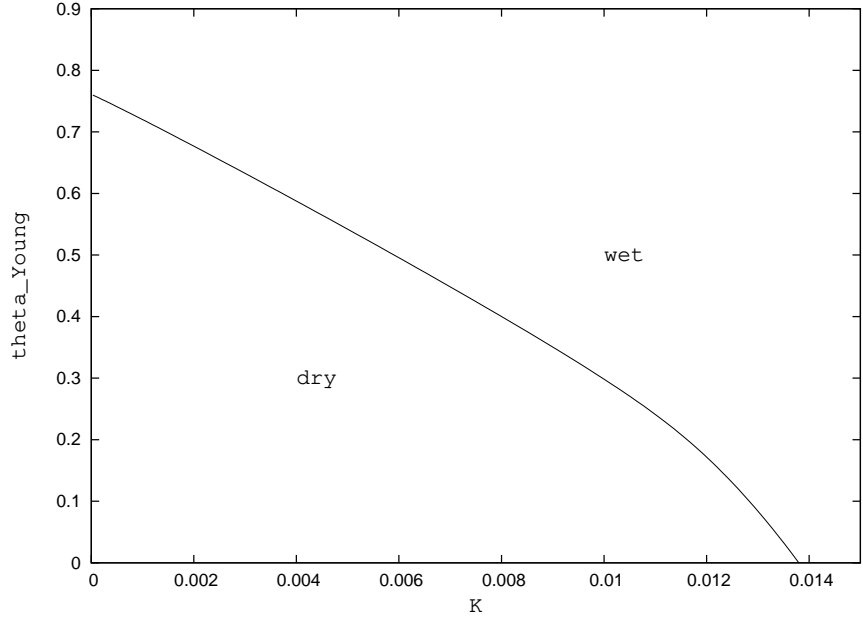}}}
\caption{The line of coexistence $F_{wet}=F_{dry}$ from (\ref{I5}, \ref{I6})
with $\sigma _{LV}=\sigma _{VS}=0.01$ and $\rho =0.2$. }
\label{iso}
\end{figure}

This leads again to an implicit critical line of coexistence in the
parameter space where $f_{dry}=f_{wet}.$ In Fig. \ref{iso}, we plot\emph{\ }%
the line of coexistence $\theta =\theta \left( K\right) $ for the arbitrary
set of parameters $\sigma _{LV}=\sigma _{VS}=0.01$ and $\rho =0.2.$ The
largest value of $\theta $ is obtained when $K=0$ as $\theta _{\max
}=2\arctan \left( \rho /0.5\right) ,$ from (\ref{I4}). The largest value of $%
K$ is obtained from the dry phase when the circle arc pinned at the corners
of the U well hits the bottom of the substrate tangentially ($\theta =0$).
We get $K_{\max }=2\rho \sigma _{LV}/\left( \rho ^{2}+0.5^{2}\right) .$

\section{SOS model: equilibrium free energies}

We now run into similar considerations for the SOS model of wetting arising
in statistical mechanics. In this Section, we compute the free energies of
the dry and wet phases in the thermodynamic limit for the SOS model (\ref{1}%
). For the sake of simplicity, we decided to work at fixed values of the
parameters $J=3.0$ and $\rho =0.2$, leaving the free parameter space be
restricted to $a$ and $K$. \newline

DRY PHASE. In the context of the SOS model (\ref{1}), the typical Wulff
shapes which come in can be described as follows. Let

\begin{equation*}
f\left( t\right) =\left( 1+J^{2}t^{2}\right) ^{1/2}-1
\end{equation*}
with derivative 
\begin{equation*}
f^{\prime }\left( t\right) =\frac{J^{2}t}{f\left( t\right) +1}.
\end{equation*}
The projected interface tension $\widetilde{\sigma }_{LV}$ corresponding to (%
\ref{1}) with $h_{i}\in \Bbb{R}$ and $a=0$ and boundary conditions implying
a slope $\tan \theta =t$, with energy measured in units of $kT,$ is defined
as 
\begin{equation}
\widetilde{\sigma }_{LV}\left( t\right) =\lim_{n\rightarrow \infty }-\frac{1%
}{n}\log Z_{n}\left( t\right) :=\widetilde{\sigma }\left( t\right) ,
\label{St}
\end{equation}
where 
\begin{equation}
\widetilde{\sigma }\left( t\right) =f\left( t\right) -\log \left( \frac{%
f\left( t\right) +2}{J}\right) .  \label{ST}
\end{equation}
We have $\widetilde{\sigma }^{\prime }\left( t\right) =\frac{J^{2}t}{f\left(
t\right) +2}.$ Then the relevant Wulff shape equations relative to (\ref{1})
and (\ref{St}) are implicitly given by (see \cite{BN}, \cite{CD} and \cite
{CDH}) 
\begin{equation*}
X\left( t\right) =\frac{n}{K}l_{0}\widetilde{\sigma }_{LV}^{\prime }\left(
t\right) ;\text{ }X\left( 0\right) =0
\end{equation*}
\begin{equation*}
Z\left( t\right) =-\frac{n}{K}l_{0}\left( \widetilde{\sigma }_{LV}\left(
t\right) -t\widetilde{\sigma }_{LV}^{\prime }\left( t\right) \right) ;\text{ 
}Z\left( 0\right) =\frac{n}{K}l_{0}\log \left( \frac{2}{J}\right)
\end{equation*}
where $t$ is the tangent $dZ/dX$ to the curve $X\rightarrow Z\left( X\right)
.$\newline

Introduce the scaled variables $x\left( t\right) =X\left( t\right) /l$ and $%
z\left( t\right) =Z\left( t\right) /l.$ Then, recalling $l=nl_{0}$%
\begin{equation}
x\left( t\right) =\frac{1}{K}\widetilde{\sigma }^{\prime }\left( t\right) ;%
\text{ }x\left( 0\right) =0  \label{W1}
\end{equation}
\begin{equation}
z\left( t\right) =-\frac{1}{K}\left( \widetilde{\sigma }\left( t\right) -t%
\widetilde{\sigma }^{\prime }\left( t\right) \right) ;\text{ }z\left(
0\right) =\frac{1}{K}\log \left( \frac{2}{J}\right)  \label{W2}
\end{equation}
where the range of $x\left( t\right) $ is $\left( -\frac{J}{K},\frac{J}{K}%
\right) $ and where $t$ is the tangent $dz/dx$ to the curve $x\rightarrow
z\left( x\right) .$ Note that $t\left( x\right) =\frac{2Kx}{J^{2}-K^{2}x^{2}}%
.$ \newline

The scaled well has now unit length and fixed height $\rho =0.2.$ As before,
fix the origin of the $\left( x,z\right) $ axis at the middle of the bottom
of the well. We wish to derive the equation of a Wulff shape (\ref{W1}, \ref
{W2}) which is pinned at $\left( x=-0.5,z=\rho \right) $ and $\left(
x=0.5,z=\rho \right) $ without hitting the bottom of the well. Let $t_{0}$
be the value of the tangent at $\left( x=0.5,z=\rho \right) $, with $%
t_{0}=t\left( 0.5\right) =\frac{K}{J^{2}-K^{2}/4}.$ We have 
\begin{equation*}
z_{dry}\left( t\right) =z\left( t\right) -z\left( t_{0}\right) +\rho
\end{equation*}
so that $z_{dry}\left( 0\right) =z\left( 0\right) -z\left( t_{0}\right) +0.2=%
\frac{1}{K}\log \left( \frac{2}{J}\right) -z\left( t_{0}\right) +\rho \geq
0. $\newline

When $K$ increases, the minimum of the hanging Wulff shape gets closer to
the bottom of the well. There is a value $K=K_{\max }$ for which this
minimum hits tangentially the bottom of the well in one point. The value of $%
K_{\max }$ is characterized by 
\begin{equation*}
z_{dry}\left( 0\right) \mid _{K=K_{\max }}=\frac{1}{K_{\max }}\log \left( 
\frac{2}{J}\right) -z\left( \frac{K_{\max }}{J^{2}-K_{\max }^{2}/4}\right)
+\rho =0.
\end{equation*}

The admissible range of $K$ in the dry regime is thus $\left[ 0,K_{\max
}\right] .$\newline

In Appendix $A2a$, we obtain the specific free energy of the dry phase as%
\emph{\ } 
\begin{equation}
f_{dry}=K\left( 0.2-z\left( t_{0}\right) \right) -2+2b\log \left[ \frac{%
1+0.5/b}{1-0.5/b}\right] .  \label{Fdry}
\end{equation}
Note that when $K\rightarrow 0$, $t_{0}\rightarrow 0$ and $z\left(
t_{0}\right) \simeq z\left( 0\right) =\frac{1}{K}\log \left( \frac{2}{J}%
\right) .$ Thus $f_{dry}\rightarrow \log \left( \frac{J}{2}\right) =%
\widetilde{\sigma }\left( 0\right) $ which is the specific free energy of
the trivial $\left( K=0\right) -$Wulff shape joining linearly the corners of
the well parallel to the bottom of the well.\newline

WET PHASE. Let $x_{1}\in \left( 0,0.5\right) $. In the wet phase, the Wulff
shape is made of two symmetric convex pieces, one joining point $\left(
x=-0.5,z=\rho \right) $ to point $\left( x=-x_{1},z=0\right) $ and the other
joining $\left( x=x_{1},z=0\right) $ to $\left( x=0.5,z=\rho \right) .$ In
between, from $\left( x=-x_{1},z=0\right) $ to $\left( x=x_{1},z=0\right) $,
the curve is flat, pinned to the substrate.

From \cite{CD}, the specific free energy of the wet part is 
\begin{equation}
\sigma _{wet}=-\log \left( \frac{2a^{2}J}{2aJ-1}\right) ,\text{ }a\geq
a_{c}:=1/J.  \label{swet}
\end{equation}
Consider the right part of the Wulff shape. Let $t_{1}$ be the tangent of
the Young angle $\theta $, which is also the interface slope at $\left(
x=x_{1},z=0\right) $; assuming $a\geq a_{c},$ we have (see (5)(8)(9) in \cite
{CD}) 
\begin{equation}
t_{1}=\tan \theta =\frac{2a\left( aJ-1\right) }{2aJ-1}\geq 0.  \label{Young}
\end{equation}
Note that when $a<a_{c}$, the affinity for the bottom of the well is not
strong enough to produce a wet part in the equilibrium Wulff shape.\newline

Let $t_{2}$ be the tangent of the angle of the Wulff shape at point $\left(
x=0.5,z=\rho \right) .$ Using the canonical equation (\ref{W1}, \ref{W2}) of
a standard Wulff shape, the equations of this Wulff shape are given by: 
\begin{eqnarray*}
x_{wet}\left( t\right) &=&x\left( t\right) -x\left( t_{2}\right) +0.5 \\
z_{wet}\left( t\right) &=&z\left( t\right) -z\left( t_{1}\right) ,
\end{eqnarray*}
where $t_{2}$ has to be determined implicitly by $z_{wet}\left( t_{2}\right)
=\rho .$ We then have 
\begin{equation*}
x_{1}=x_{wet}\left( t_{1}\right) =x\left( t_{1}\right) -x\left( t_{2}\right)
+0.5.
\end{equation*}
\newline
The specific free energy of the wet phase is obtained as (see Appendix $A2b$%
) 
\begin{equation}
f_{wet}=2x_{1}\sigma _{wet}-2\left( 0.5-x_{1}\right) \left( Kz\left(
t_{1}\right) +2\right) +2b\log \frac{\left( 1+\frac{x\left( t_{2}\right) }{b}%
\right) \left( 1-\frac{x_{1}+x\left( t_{2}\right) -0.5}{b}\right) }{\left( 1+%
\frac{x_{1}+x\left( t_{2}\right) -0.5}{b}\right) \left( 1-\frac{x\left(
t_{2}\right) }{b}\right) }.  \label{Fwet}
\end{equation}
\newline

When $K=0$, the two pieces of the Wulff shapes become straight lines. The
tangent of the Young angle is thus $t_{1}=\rho /\left( 0.5-x_{1}\right) $
and 
\begin{equation}
f_{wet}\mid _{K=0}=2\left( 0.5-x_{1}\right) \widetilde{\sigma }\left(
t_{1}\right) +2x_{1}\sigma _{wet}.  \label{FwetK0}
\end{equation}
There exists a maximal value $a_{\max }$ of $a$ characterized by: $%
f_{wet}\mid _{K=0}=f_{dry}\mid _{K=0}=\log \left( J/2\right) .$ For $%
a>a_{\max }$, the wet phase has a lower free energy than the dry phase, for
all $K\ge 0$.

\begin{figure}[tbp]
\centerline{\ \mbox{\includegraphics[width=12cm]{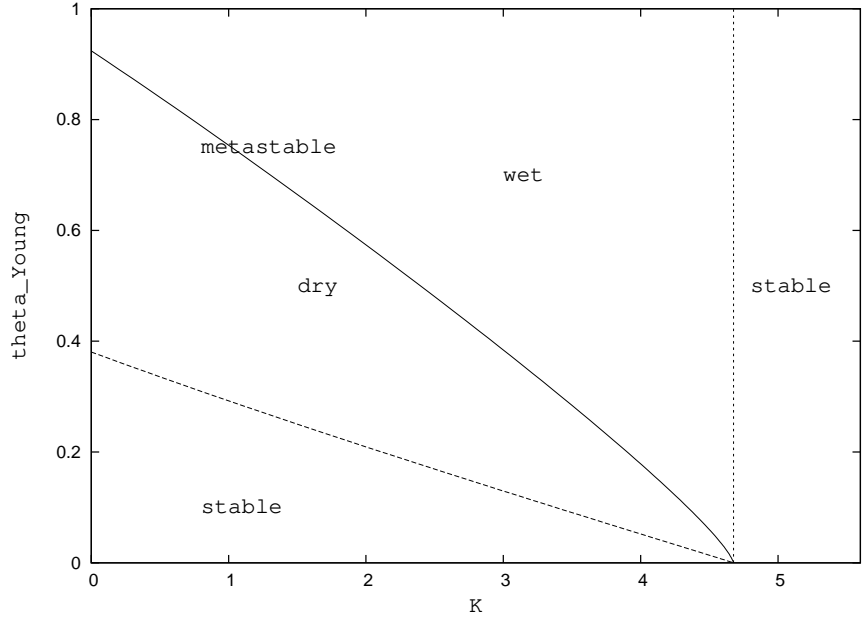}}}
\caption{ The line of coexistence $f_{dry}=f_{wet}$ from (\ref{Fdry}, \ref
{Fwet}), together with the metastability lines: $f_{dry}=f_{sad}$ from (\ref
{Fdry}, \ref{Fsad}) and $f_{wet}=f_{sad}$ from (\ref{Fwet}, \ref{Fsad});
here $J=3.0$ and $\rho =0.2$. }
\label{theKcol}
\end{figure}

We fix $J=3.0$ and look for the values $a=a\left( K\right) $ for which $%
f_{wet}=f_{dry}$, using (\ref{Fdry}) and (\ref{Fwet}), meaning coexistence
of the two phases. In this example, the range of $K$ is $\left[ 0,K_{\max
}\right] $ with $K_{\max }\simeq 4.6767$ and the range of $a$ is $\left[
a_{c},a_{\max }\right] $ with $a_{c}=1/J$ and $a_{\max }\simeq 1.5114$.
Using this curve $a=a\left( K\right) $, together with (\ref{Young}),
relating the Young angle $\theta $ to $a$, we rather consider the line of
coexistence $\theta =\theta \left( K\right) .$ This line of coexistence is
shown on Fig. \ref{theKcol}. In this phase diagram plot, the dotted lines
separate a metastable region where the two phases coexist and stable regions
where a single phase (either wet or dry) is stable; the solid line of
coexistence separates the two stable phases within the metastable region.
The two dotted lines are obtained while using $f_{sad}=f_{dry}$ and $%
f_{sad}=f_{wet},$ respectively. Note that the point at $K=0$ separating the
dry stable zone from the metastable zone is exactly characterized by $\rho
=0.2=0.5\tan \theta $ as in (\ref{I2}).

\section{Dynamics and numerical simulations}

Let us first compute the equilibrium free energy of the saddle-point phase,
characterized by a single contact point in the center of the well.\newline

SADDLE-POINT PHASE. In the saddle-point phase, the Wulff shape is made of
two symmetric pieces and $x_{1}=0$ so that there is no flat part
corresponding to wetting. With $t_{1}$ the tangent of the Young angle at
point $\left( x=0,z=0\right) $, the equations of the Wulff shape in the
saddle-point configuration are 
\begin{eqnarray*}
x_{sad}\left( t\right) &=&x\left( t\right) -x\left( t_{1}\right) \\
z_{sad}\left( t\right) &=&z\left( t\right) -z\left( t_{1}\right) .
\end{eqnarray*}
In Appendix $A2c$, we show that the specific free energy of the saddle phase
reads 
\begin{equation}
f_{sad}=-Kz\left( t_{1}\right) -2+2b\log \frac{\left( 1+\frac{0.5+x\left(
t_{1}\right) }{b}\right) \left( 1-\frac{x\left( t_{1}\right) }{b}\right) }{%
\left( 1+\frac{x\left( t_{1}\right) }{b}\right) \left( 1-\frac{0.5+x\left(
t_{1}\right) }{b}\right) },  \label{Fsad}
\end{equation}
which is implicitly known because so is $t_{1}.$

\subsection{Dynamics\textbf{.}}

For a square well of width $l=nl_{0}$, we consider a Markovian dynamics of
Monte-Carlo type having (\ref{1}) as invariant measure. The free energy
barrier to cross starting from the dry (wet) phase is $%
F_{sad}^{(n)}-F_{dry}^{(n)}\sim nl_{0}(f_{sad}-f_{dry})$ (respectively $%
F_{sad}^{(n)}-F_{wet}^{(n)}\sim nl_{0}(f_{sad}-f_{wet})$). Conventional
wisdom suggests that, with $\left\langle \tau _{dw}^{\left( n\right)
}\right\rangle $ the mean time needed to first enter the wet (dry) phase
starting from the dry (wet) phase, in a system of size $l=nl_{0}$ as $n\to
\infty $, 
\begin{eqnarray*}
\frac{1}{nl_{0}}\log \left\langle \tau _{dw}^{\left( n\right) }\right\rangle
&\rightarrow &f_{sad}-f_{dry} \\
\frac{1}{nl_{0}}\log \left\langle \tau _{wd}^{\left( n\right) }\right\rangle
&\rightarrow &f_{sad}-f_{wet}.
\end{eqnarray*}
The limiting quantities coincide, assuming coexistence, $\theta =\theta (K)$%
. We expect polynomial corrections such as $e^{-nl_{0}\left(
f_{sad}-f_{dry}\right) }\left\langle \tau _{dw}^{\left( n\right)
}\right\rangle \sim c_{dw}n^{2}$ and $e^{-nl_{0}\left(
f_{sad}-f_{wet}\right) }\left\langle \tau _{wd}^{\left( n\right)
}\right\rangle \sim c_{wd}n^{2}$ where $c_{dw}$ and $c_{wd}$ would be two
possibly distinct $n-$independent constants. In any case, the relative
weights of the dry and wet phases may differ, 
\begin{equation}
\lim_{n\to \infty }\frac{\left\langle \tau _{dw}^{\left( n\right)
}\right\rangle }{\left\langle \tau _{dw}^{\left( n\right) }\right\rangle
+\left\langle \tau _{wd}^{\left( n\right) }\right\rangle }\ne \lim_{n\to
\infty }\frac{\left\langle \tau _{wd}^{\left( n\right) }\right\rangle }{%
\left\langle \tau _{dw}^{\left( n\right) }\right\rangle +\left\langle \tau
_{wd}^{\left( n\right) }\right\rangle }\,.  \label{lim}
\end{equation}


\subsection{Numerical simulation}

Simulations were performed for one-dimensional interfaces over a trough of
length $l=n$ and depth $h=n/5$. The interface is pinned at both ends, $%
h_{0}=h_{n+1}=n/5$, and $\mathbf{h}=(h_{1},\dots ,h_{n})\in R_{+}^{n}$ is
distributed at equilibrium according to (\ref{1}), or 
\begin{equation}
\mu _{n}(d\mathbf{h})=Z_{n}^{-1}\exp (-J\sum_{i=0}^{n}|h_{i+1}-h_{i}|-\frac{K%
}{n}\sum_{i=1}^{n}h_{i})\prod_{i=1}^{n}(1+a\delta (h_{i}))dh_{i}  \label{mun}
\end{equation}
where the partition function $Z_{n}$ normalizes the probability.

\begin{figure}[tbp]
\centerline{\ \mbox{\includegraphics[width=10cm]{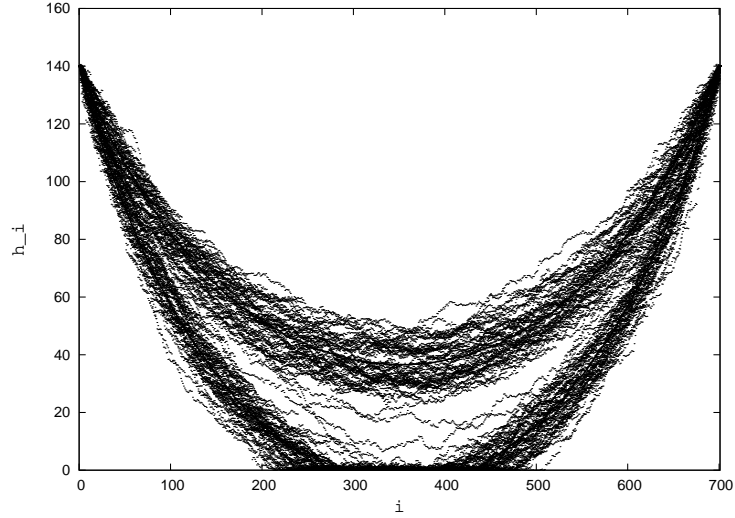}} 
}
\caption{One hundred samples of the interface for $n=700,J=3.0,K=4.0,a=0.455$
(dotted lines) and the two wulff shapes (solid lines).}
\label{Figh1}
\end{figure}

The sub-lattice parallel heat bath dynamics, irreducible and satisfying the
detailed balance condition with respect to $\mu _{n}$, is defined as
follows, for $t=1,2,\dots $: 
\begin{equation}
P(d\mathbf{h}^{t}|\mathbf{h}^{t-1})=\mu _{n}(d\mathbf{h}^{t})\prod_{i+t\ 
\mathrm{odd}}\delta (h_{i}^{t}-h_{i}^{t-1})/\int \mu _{n}(d\mathbf{h}%
)\prod_{i+t\ \mathrm{odd}}\delta (h_{i}-h_{i}^{t-1})  \label{oe}
\end{equation}
Fig. \ref{Figh1} shows one hundred samples obtained from this dynamics. The
interface is typically near one of the two Wulff shapes, ``dry'' or ``wet''.
The parameters $J,K,a$ were chosen so that the free energies computed from
the two Wulff shapes were approximately equal, so that the interface spent
approximately equal times near these two shapes. The empty region between
the two shapes indicates a region of small probability, which we shall call
a free energy barrier. 
The aim of the simulation is to find the law of the escape time from dry to
wet or conversely. The interface can be within typical fluctuation from one
of the Wulff shapes, in which case it is easy to decide whether it is
``dry'' or ``wet'', but large deviations in between cannot be attached to
one or the other in any justified way.

We measure time either in Monte-Carlo Steps per Site (MCS/S), or in
Monte-Carlo Steps per Site divided by $n^{2}$ (MCS/S/$n^{2}$), because the
relaxation time of a flat interface without free energy barrier is of order $%
n^{2}$ MCS/S in non-conservative dynamics. One MCS/S corresponds to two time
steps of the form (\ref{oe}). This observation may well have interesting
consequences to analyze the dynamics of spreading of nanodrops on top of
rough substrates using molecular dynamics simulations.

\begin{figure}[tbp]
\centerline{\ \mbox{\includegraphics[width=13cm]{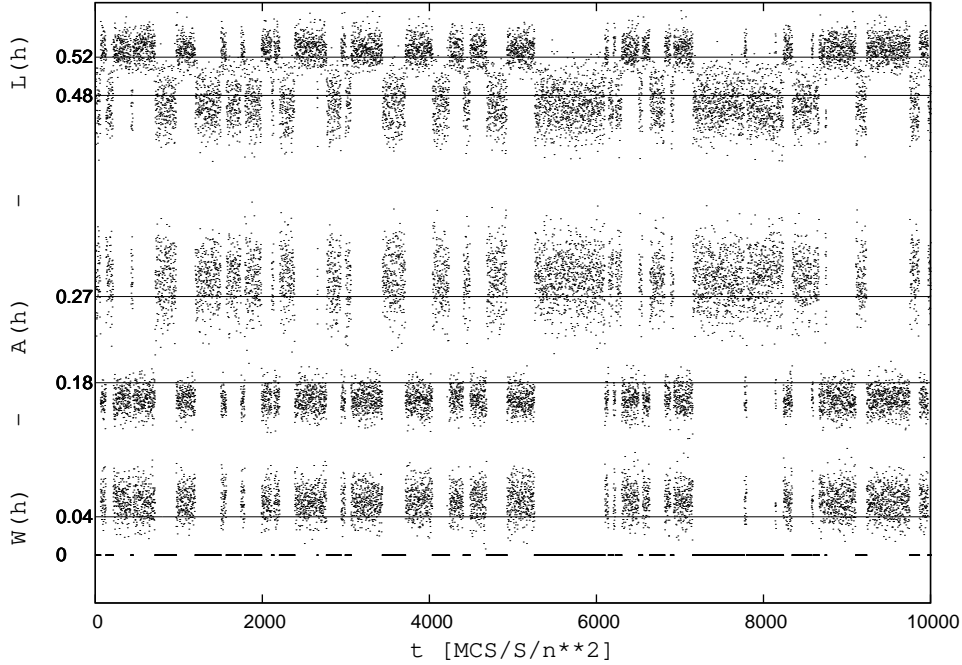}} 
}
\caption{From top to bottom: $L(\mathbf{h})$, $A(\mathbf{h})$, $W(\mathbf{h}%
) $, function of time measured in MCS/S/$n^{2}$ (dots), with $%
n=700,J=3.0,K=4.0,a=0.455$. The regions between 0.48 and 0.52 for $L(\mathbf{%
h})$, between 0.18 and 0.27 for $A(\mathbf{h})$, or between 0 and 0.04 for $%
W(\mathbf{h})$, correspond to the barrier between `dry' and `wet'.}
\label{jka}
\end{figure}

At time intervals of the order of a few MCS/S, three measurements are taken,
corresponding to the observables that make up the Hamiltonian in (\ref{mun}%
): the normalized length of the interface 
\begin{equation}
L(\mathbf{h})=\frac{1}{n+1}\sum_{i=0}^{n}|h_{i+1}-h_{i}|\,,  \label{L}
\end{equation}
the normalized area below the interface 
\begin{equation}
A(\mathbf{h})=\frac{3}{n^{2}}\sum_{i=1}^{n}h_{i}\,,  \label{A}
\end{equation}
and the number of zeroes of $\mathbf{h}$ divided by $2n$, denoted $W(\mathbf{%
h})$ (number of contacts with Wall). 
Each measurement appears as a dot on Fig. \ref{jka}, making up clouds of
points around the corresponding mean values for each of the two Wulff
shapes. The factors of 2 and 3 in the definitions of $W(\mathbf{h})$ and $A(%
\mathbf{h})$ are designed to facilitate the reading in Fig. \ref{jka}. 
We then decide somewhat arbitrarily intervals of values of the three
observables associated to the wet state or to the dry state or to the region
in between, called the barrier: 
\begin{equation*}
\begin{array}{cccc}
\mathrm{Dry:}\quad & L(\mathbf{h})<0.48\,,\quad & A(\mathbf{h})>0.27\,,\quad
& W(\mathbf{h})=0 \\ 
\mathrm{Wet:}\quad & L(\mathbf{h})>0.52\,,\quad & A(\mathbf{h})<0.18\,,\quad
& W(\mathbf{h})>0.04 \\ 
\mathrm{Barrier:}\quad & \mathrm{otherwise} &  & 
\end{array}
\end{equation*}
At any given time, if all three observables agree for dry, or for wet, then
the interface is declared dry (D), or wet (W). Otherwise it is considered to
be in the barrier (B) between dry and wet. The barrier is clearly in a
region of large deviations from the dry state or from the wet state. In
principle one observable should be enough, but the numerical simulation is
done with a limited set of values of $n$, and the picture emerges more
clearly using three observables.

We thus obtain a marginal of the interface dynamics, which is a
non-Markovian process with values in \{D,B,W\}. Of course the simulation
uses discrete time, but the mean sojourn time in D or B or W is of order $%
n^2 $ MCS/S, so that a continuous time description is better, with numerical
results expressed in MCS/S/$n^2$.

A marginal of this marginal is the sequence of letters, forgetting their
duration, looking like

DBDBDBD\textbf{B}WBWBWBWBWBWBWBW\textbf{B}DBDBDBDBD\textbf{B}WBWB\dots

with only four 2-letter patterns: DB, BD, WB, BW. This restriction allows
only six 3-letter patterns: DBD, BDB, WBW, BWB, DBW, WBD. The last two are
very rare ($\sim \exp (-\mathrm{const.}n)$). The numerical algorithm does
not strictly forbid the 2-letter patterns DW or WD, crossing the barrier in
a few MCS/S, but they are so rare, with a relative frequency expected $\sim
\exp (-\mathrm{const.}n^{2})$, that we haven't seen them in the experiment.

The limiting relative frequencies of the letters D, B, W, are respectively
1/4, 1/2, 1/4. The barrier B should be decomposed into a dry side and a wet
side of the saddle point. We cannot describe precisely the corresponding
configurations, but we can assume that within a 3-letter pattern DBD, the
system remains in the dry state, or on the dry side of the saddle point, and
analogously for WBW. If we would change every B in DBD into D, and every B
in WBW into W, then we would find relative frequencies tending to 1/2, 1/2
for D and W. The corresponding partition of the configuration space into D
and W is analogous to partitions which play a role in some studies of
metastability \cite{BEGK}. There is of course a remainder $\sim\exp(-\mathrm{%
const.}n)$ for \textbf{B} from D\textbf{B}W and W\textbf{B}D.

The dynamics is started at time $t_{0}=0$ in D, with the interface near the
Wulff shape associated with the dry state. The first time in W is denoted $%
t_{1}$, measured in MCS/S/$n^{2}$. The next time in D is denoted $t_{2}$,
etc. Thus $t_{2k+1}$ is the first time in W after $t_{2k}$ for $k=0,1,\dots
,k_{\max }$ and $t_{2k}$ is the first time in D after $t_{2k-1}$ for $%
k=1,2,\dots ,k_{\max }$. We also define $t_{2k}^{\prime }$ as the last time
in D before $t_{2k+1}$ and after $t_{2k}$, and $t_{2k+1}^{\prime }$ as the
last time in W before $t_{2k+2}$ and after $t_{2k+1}$. The empirical mean
escape times from D and from W are defined respectively as 
\begin{equation}
\bar{\tau}_{D}=\frac{1}{k_{\max }}\sum_{k=0}^{k_{\max }-1}(t_{2k}^{\prime
}-t_{2k})\,,\text{ }\bar{\tau}_{W}=\frac{1}{k_{\max }}\sum_{k=0}^{k_{\max
}-1}(t_{2k+1}^{\prime }-t_{2k+1})  \label{td}
\end{equation}
and the empirical mean barrier crossing time is defined as 
\begin{equation}
\bar{\tau}_{B}=\frac{1}{2k_{\max }}\sum_{\ell =0}^{2k_{\max }-1}(t_{\ell
+1}-t_{\ell }^{\prime })  \label{tb}
\end{equation}
The total time of the experiment is $k_{\max }(\bar{\tau}_{D}+\bar{\tau}%
_{W})+2k_{\max }\bar{\tau}_{B}$, in MCS/S/$n^{2}$. The measured escape times
from D and from W are respectively $\tau _{k}^{D}=t_{2k}^{\prime }-t_{2k}$
and $\tau _{k}^{W}=t_{2k+1}^{\prime }-t_{2k+1}$. They are not a priori
independent nor identically distributed, but one may study the empirical
cumulative distribution function of a random variable $\tau ^{D}$ or $\tau
^{W}$ having the sequence of realizations $\tau _{k}^{D}$ or $\tau _{k}^{W}$%
: 
\begin{equation}
R^{D}(t)=\frac{1}{k_{\max }}|\{k:\tau _{k}^{D}\le t\}|\,,\qquad R^{W}(t)=%
\frac{1}{k_{\max }}|\{k:\tau _{k}^{W}\le t\}|  \label{RD}
\end{equation}
\begin{figure}[tbp]
\centerline{\mbox{\includegraphics[width=13cm]{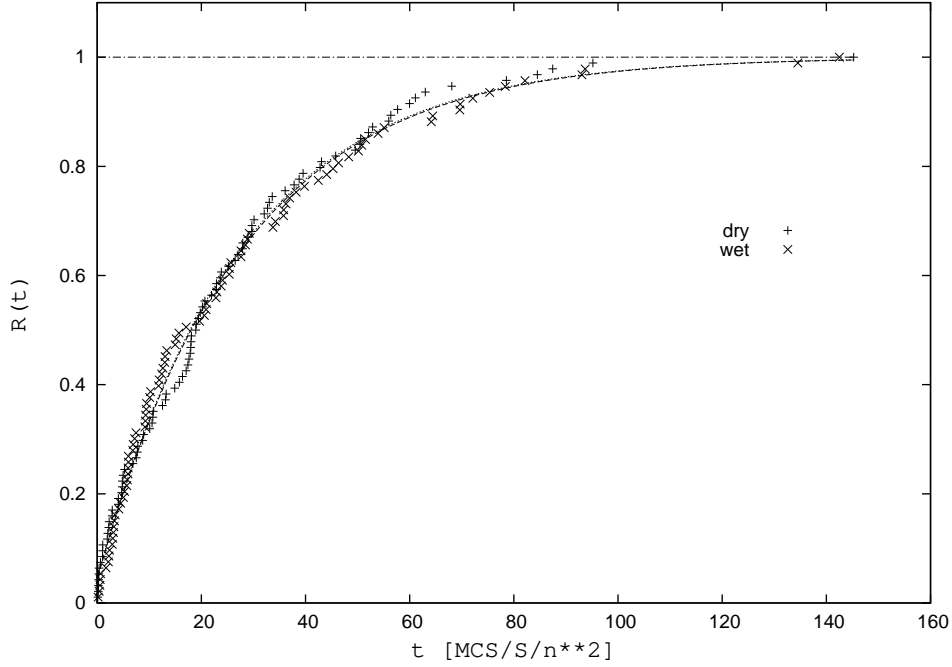}}}
\caption{Empirical cumulative distribution function of $\tau _{k}^{D}$ (dry)
and $\tau _{k}^{W}$ (wet) and fit $R^{D}(t)=1-\exp (-(t+1.2)/27.8)$, $%
R^{W}(t)=1-\exp (-(t+1.3)/27.5)$, functions of time in MCS/S/$n^{2}$, for $%
n=500,J=3.0,K=4.0,a=0.451$, and $k_{\max }=93$.}
\label{tau}
\end{figure}
This is shown on Fig. \ref{tau}, showing that $\tau ^{D}$ and $\tau ^{W}$
are approximately independent and identically distributed (iid) exponential
random variables. The fit is better with a small shift as indicated, which
we interpret as a redistribution of the barrier time $2k_{\max }\bar{\tau}%
_{B}$ to the dry and wet states, in a proportion which we cannot decide
directly.

\begin{figure}[tbp]
\centerline{\mbox{\includegraphics[width=13cm]{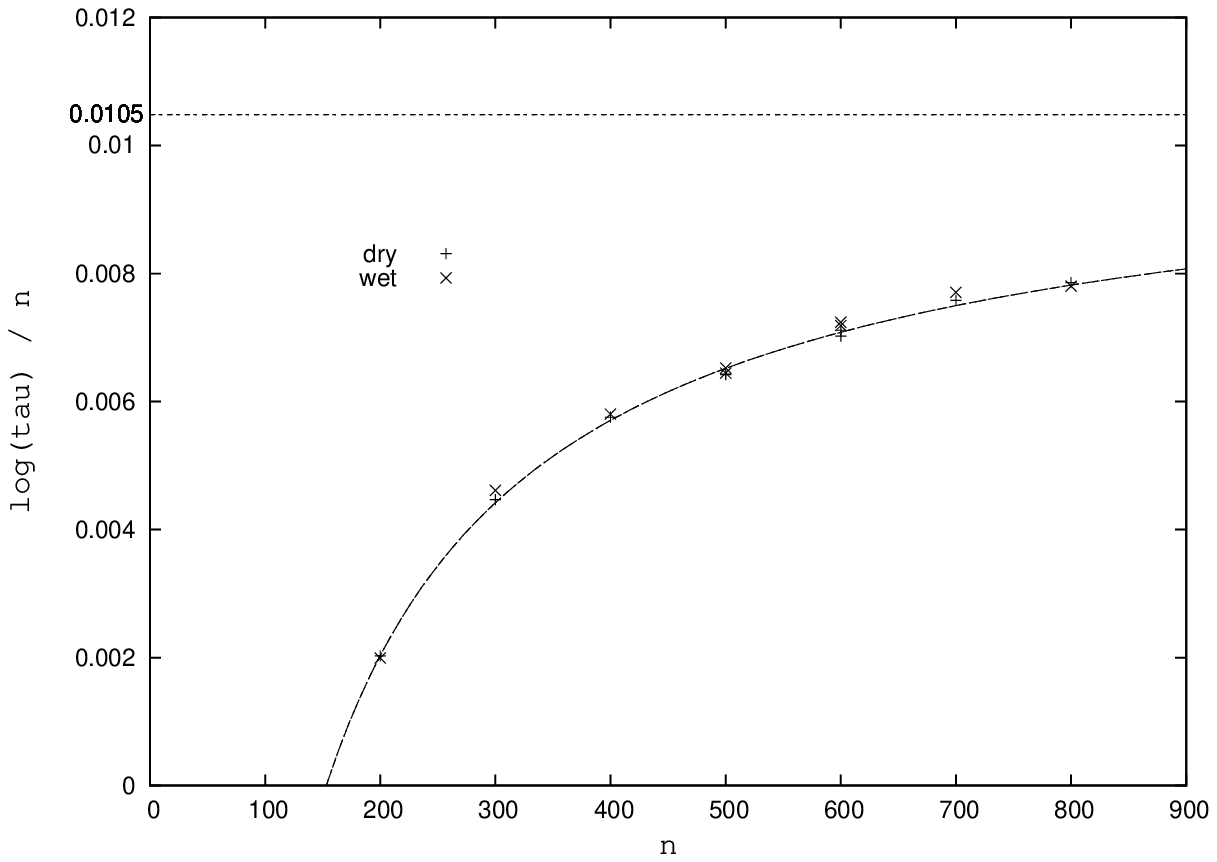}}}
\caption{$\log(\bar\tau^D)/n$ (dry) and $\log(\bar\tau^W)/n$ (wet) and fit $%
c+d*\log(n)/n$ for $J=3.0,K=4.0$, and $a=0.4375,\,0.445,\,0.449,\,0.451,%
\,0.453,\,0.455,\,0.456$ for $n=200,300,400,500,600,700,800$ respectively.}
\label{dbw}
\end{figure}
We now turn to the dependence of $\bar\tau^B$, $\bar\tau^D$ and $\bar\tau^W$
upon $n$. Fig. \ref{dbw} shows that $\log(\bar\tau^D)/n$ and $%
\log(\bar\tau^W)/n$ tend to a constant $\simeq 0.0105$ as $n\to\infty$, to
be compared with the free energy density difference between the barrier or
saddle point and the dry or wet states, giving a theoretical value $0.0129$,
cf. (\ref{Fsad})(\ref{Fdry})(\ref{Fwet}). On the other hand $\bar\tau^B$,
measured like the other times in MCS/S/$n^2$ appears to converge to a limit
independent of $n$, $\bar\tau^B=0.22\pm0.03$ for $n=500, 600, 700, 800$.

For each value of $n$, the simulations were done with a value $a_n$ of the
parameter $a$ chosen so that the system spent equal time in `dry' and in
`wet'. Finite size corrections to the thermodynamic limit should imply $%
a_n=a+O(1/n)$; a fit by $a_n=a_\infty+\mathrm{const.}/n$ gives $%
a_\infty=4.61 $, with an uncertainty compatible with the expected limit $%
a\simeq4.64$, as can be read on Fig. \ref{theKcol} at $K=4.0$.

The size of samples used in Fig. \ref{dbw} for $%
n=200,300,400,500,600,700,800 $ was $k_{\max }=15702,4936,2433,778,272,47,19$
respectively, where $k_{\max }$ is the number of observed transitions from
`dry' to `wet', also equal to the number of observed transitions from `wet'
to `dry'.

\section{Conclusion}

We describe a situation where a liquid droplet is on top of a structured
substrate presenting grooves or wells. We show that there exists a range of
parameters for which wetting and non-wetting states both share the same free
energy, entailing that there can be transitions between these two states. We
calculate analytically the involved free energies of the dry and wet states
in the context first of a simple thermodynamical model and then for a random
microscopic model based on the Solid-on-Solid (SOS) model. In the latter
case, we present a dynamical model showing stochastic transitions between
the wetting and non-wetting states. We show that the expected time to switch
from one state to the other grows exponentially with the free energy barrier
between the stable states and the saddle state, proportional to the width of
the grooves. This random time is shown to have an exponential-like
distribution. Our study hopefully contributes to a better comprehension of
the behavior of fluids on structured surfaces.

\section{Appendix}

A1. Proof of (\ref{I5}).\newline

Let $O$\ be the center of the circle anchored at $A$\ and $B$\ with scaled
radius $R$. With $\varphi $\ the angle $\left( OA,OB\right) ,$\ the scaled
euclidean distance between $A$\ and $B$\ is 
\begin{equation*}
AB=2R\sin \left( \frac{\varphi }{2}\right) =\sqrt{\left( 0.5-x_{1}\right)
^{2}+\rho ^{2}}.
\end{equation*}
The scaled arc length of the arc joining $A$\ to $B$\ is thus 
\begin{equation*}
\widehat{AB}=R\varphi =2R\arcsin \left( \frac{\sqrt{\left( 0.5-x_{1}\right)
^{2}+\rho ^{2}}}{2R}\right) .
\end{equation*}
Using $\sin \left( \theta +\varphi /2\right) =\frac{\rho }{AB}$\ and $%
AB=2R\sin \left( \frac{\varphi }{2}\right) $\ gives $AB,$\ after eliminating 
$\varphi .$\ With $\Delta =\left( \sin ^{2}\theta +\frac{\rho \cos \theta }{R%
}\right) ^{2}-\left( \frac{\rho }{R}\right) ^{2},$\ we get $AB$\ as the
explicit function of $\left( \theta ,\rho ,R\right) :$%
\begin{equation*}
AB^{2}=2R^{2}\left( \sin ^{2}\theta +\frac{\rho \cos \theta }{R}-\sqrt{%
\Delta }\right) .
\end{equation*}
Note that, geometrically, $\rho <R\cos \theta $\ which entails that $\Delta
>0.$

As a result, $\widehat{AB}=2R\arcsin \left( AB/\left( 2R\right) \right) ,$\ $%
x_{1}=0.5-\sqrt{AB^{2}-\rho ^{2}}$\ and $\varphi =2\arcsin \left( AB/\left(
2R\right) \right) $\ have themselves an explicit expression in terms of $%
\left( \theta ,\rho ,R\right) $. These quantities are the ones needed to
compute $f_{wet}$\ from (\ref{I5}).\newline

A2. Proofs of (\ref{Fdry}), (\ref{Fwet}) and (\ref{Fsad}).\newline

$A2a$\ (DRY PHASE). The specific free energy $f_{dry}\left(
x_{1},x_{2}\right) :=\lim_{l\rightarrow \infty }\frac{F_{dry}\left(
x_{1},x_{2}\right) }{l}$\ of the dry phase between $\left(
x_{1},x_{2}\right) $\ has two contributions, one pertaining to the length
and the other to the area below the interface, namely: 
\begin{equation}
f_{dry}\left( x_{1},x_{2}\right) =\int_{x_{1}}^{x_{2}}\widetilde{\sigma }%
dx+K\int_{x_{1}}^{x_{2}}z_{dry}dx=\int_{x_{1}}^{x_{2}}t\widetilde{\sigma }%
^{\prime }dx+K\left( x_{2}-x_{1}\right) \left( 0.2-z\left( t_{0}\right)
\right) .  \label{Fdry1}
\end{equation}
With $b:=J/K,$\ we have 
\begin{equation*}
\int_{x_{1}}^{x_{2}}t\widetilde{\sigma }^{\prime
}dx=K\int_{x_{1}}^{x_{2}}txdx=\int_{x_{1}}^{x_{2}}\frac{2K^{2}x^{2}}{%
J^{2}-K^{2}x^{2}}dx=2\int_{x_{1}}^{x_{2}}\frac{x^{2}/b^{2}}{1-x^{2}/b^{2}}dx
\end{equation*}
\begin{equation*}
=-2\left( x_{2}-x_{1}\right) +\int_{x_{1}}^{x_{2}}\left( \frac{1}{1-x/b}+%
\frac{1}{1+x/b}\right) dx
\end{equation*}
\begin{equation*}
=-2\left( x_{2}-x_{1}\right) +b\log \left[ \frac{1+x_{2}/b}{1+x_{1}/b}\frac{%
1-x_{1}/b}{1-x_{2}/b}\right] .
\end{equation*}
Finally, letting $x_{1}=-0.5,x_{2}=0.5$, we obtain the specific free energy
of the dry phase as in (\ref{Fdry}).\newline

$A2b$\ (WET PHASE). The limiting specific free energy of the wet phase has
two parts, one corresponding to the symmetric pieces of the Wulff shape, the
other to the flat part: 
\begin{equation}
f_{wet}=2\int_{x_{1}}^{0.5}\left( \widetilde{\sigma }+Kz_{wet}\right)
dx_{wet}+2x_{1}\sigma _{wet}.  \label{Fwet1}
\end{equation}
We have 
\begin{equation*}
\int_{x_{1}}^{x_{2}}\left( \widetilde{\sigma }+Kz_{wet}\right)
dx_{wet}=-Kz\left( t_{1}\right) \left( x_{2}-x_{1}\right)
+\int_{x_{1}+x\left( t_{2}\right) -0.5}^{x_{2}+x\left( t_{2}\right)
-0.5}\left( \widetilde{\sigma }+Kz\right) dx
\end{equation*}
and, using 
\begin{equation*}
\int_{\alpha }^{\beta }\left( \widetilde{\sigma }+Kz\right) dx=\int_{\alpha
}^{\beta }t\widetilde{\sigma }^{\prime }dx=-2\left( \beta -\alpha \right)
+b\log \frac{\left( 1+\frac{\beta }{b}\right) \left( 1-\frac{\alpha }{b}%
\right) }{\left( 1+\frac{\alpha }{b}\right) \left( 1-\frac{\beta }{b}\right) 
},
\end{equation*}
we finally obtain 
\begin{equation*}
f_{wet}=2x_{1}\sigma _{wet}-2Kz\left( t_{1}\right) \left( 0.5-x_{1}\right)
+2\int_{x_{1}+x\left( t_{2}\right) -0.5}^{x\left( t_{2}\right) }\left( 
\widetilde{\sigma }+Kz\right) dx,
\end{equation*}
leading to (\ref{Fwet}).\newline

$A2c$\ (SADDLE PHASE). The slope $t_{2}$\ of the interface at point $\left(
x=0.5,z=\rho \right) $\ has to be determined implicitly by $z_{sad}\left(
t_{2}\right) =\rho .$\ We have 
\begin{equation*}
x_{sad}\left( t_{2}\right) =x\left( t_{2}\right) -x\left( t_{1}\right) =0.5
\end{equation*}
\begin{equation*}
z_{sad}\left( t_{2}\right) =z\left( t_{2}\right) -z\left( t_{1}\right) =\rho
.
\end{equation*}
From the first equation, recalling $x\left( t\right) =\frac{1}{K}$\ $%
\widetilde{\sigma }^{\prime }\left( t\right) $\ and $\widetilde{\sigma }%
^{\prime }\left( t\right) =\frac{J^{2}t}{f\left( t\right) +2}=\frac{J^{2}t}{%
\left( 1+J^{2}t^{2}\right) ^{1/2}+1}$\ 
\begin{equation*}
t_{2}=\widetilde{\sigma }^{\prime -1}\left( \widetilde{\sigma }^{\prime
}\left( t_{1}\right) +0.5K\right) =:\varphi \left( t_{1}\right)
\end{equation*}
where the inverse of $\widetilde{\sigma }^{\prime }$\ is easily seen to be 
\begin{equation*}
\widetilde{\sigma }^{\prime -1}\left( s\right) =\frac{2s}{J^{2}-s^{2}}.
\end{equation*}
Thus $t_{2}=\varphi \left( t_{1}\right) $\ is an explicit known function of $%
t_{1}.$\ Plugging this expression in the second equation and recalling $%
z\left( t\right) =-\left( \widetilde{\sigma }\left( t\right) -t\widetilde{%
\sigma }^{\prime }\left( t\right) \right) /K$, 
\begin{equation*}
\left( \widetilde{\sigma }-t\widetilde{\sigma }^{\prime }\right) \left(
\varphi \left( t_{1}\right) \right) -\left( \widetilde{\sigma }-t\widetilde{%
\sigma }^{\prime }\right) \left( t_{1}\right) =-\rho K
\end{equation*}
giving $t_{1}$\ implicitly and then $t_{2}$\ using $t_{2}=\varphi \left(
t_{1}\right) $. The specific free energy of the saddle-point phase is thus 
\begin{equation}
f_{sad}=2\int_{0}^{0.5}\left( \widetilde{\sigma }+Kz_{sad}\right) dx_{sad}.
\label{Fsad1}
\end{equation}
We have 
\begin{equation*}
\int_{x_{1}}^{x_{2}}\left( \widetilde{\sigma }+Kz_{sad}\right)
dx_{sad}=-Kz\left( t_{1}\right) \left( x_{2}-x_{1}\right)
+\int_{x_{1}+x\left( t_{1}\right) }^{x_{2}+x\left( t_{1}\right) }\left( 
\widetilde{\sigma }+Kz\right) dx
\end{equation*}
and so 
\begin{equation*}
f_{sad}=-Kz\left( t_{1}\right) +2\int_{x\left( t_{1}\right) }^{0.5+x\left(
t_{1}\right) }\left( \widetilde{\sigma }+Kz\right) dx,
\end{equation*}
leading to (\ref{Fsad}).\newline

B. A toy model.\newline

Although the problems encountered in this study are far from being
Markovian, we find it useful to end up with recalling similar issues in the
context of Markov chains or the like.\newline

Consider a discrete-time $k$\ Markov chain $X_{k}$\ with five states $%
\left\{ 0,1,2,3,4\right\} .$\ Suppose the following transition probabilities 
$P_{i,j}$\ from state $i$\ to $j$\ hold: $P_{0,0}=1-\alpha ^{\prime }$, $%
P_{0,1}=\alpha ^{\prime }$; $P_{1,0}=1-\alpha $, $P_{1,2}=\alpha $; $%
P_{2,1}=1/2$, $P_{2,3}=1/2$; $P_{3,2}=\alpha $, $P_{3,4}=1-\alpha $; $%
P_{4,3}=\alpha ^{\prime }$, $P_{4,4}=1-\alpha ^{\prime }.$

The parameters $\alpha $\ and $\alpha ^{\prime }$\ are small, with 
\begin{eqnarray*}
\alpha &=&e^{-\left[ U\left( 2\right) -U\left( 3\right) \right] /\varepsilon
}=e^{-\left[ U\left( 2\right) -U\left( 1\right) \right] /\varepsilon } \\
\alpha ^{\prime } &=&e^{-\left[ U\left( 3\right) -U\left( 4\right) \right]
/\varepsilon }=e^{-\left[ U\left( 1\right) -U\left( 0\right) \right]
/\varepsilon },
\end{eqnarray*}
the energy barrier terms within the brackets being all positive and $%
\varepsilon $\ small. Thus $\left\{ 0\right\} $\ and $\left\{ 4\right\} $\
are two stable states separated by a barrier state $\left\{ 2\right\} $. Let
us compute the law of the time $\tau _{0,4}=\inf \left( k:X_{k}=4\mid
X_{0}=0\right) $\ needed to move from state $\left\{ 0\right\} $\ to state $%
\left\{ 4\right\} .$\ The chain is a nearest neighbors birth and death chain
which is ergodic. Putting $p_{x}=P_{x,x+1}$\ and $q_{x}=P_{x,x-1}$, the
invariant measure is $\pi _{x}=\pi _{0}\prod_{y=0}^{x-1}\frac{p_{y}}{q_{y+1}}
$. Starting from $\left\{ 0\right\} $, the sample paths are made of iid
excursions separating consecutive visits to $\left\{ 0\right\} .$\ The law
of the height $H$\ of an excursion is given by 
\begin{equation*}
\Pr \left( H\geq h\right) =\frac{1}{\varphi \left( h\right) },
\end{equation*}
where 
\begin{equation*}
\varphi \left( x\right) =1+\sum_{y=1}^{x-1}\prod_{z=1}^{y}\frac{q_{z}}{p_{z}}
\end{equation*}
is the scale function of the chain. In particular, we get $\Pr \left(
H=4\right) =1/\varphi \left( 4\right) =\alpha /2.$

With $\mu $\ the mean length of an excursion and $H_{i}$\ the height of
excursion $i$, we have 
\begin{equation*}
\Pr \left( \tau _{0,4}>K\right) =\Pr \left( \sup_{k\leq K}X_{k}<4\right)
\approx \Pr \left( \max_{i=1,..,\left[ K/\mu \right] }H_{i}<4\right) =\Pr
\left( H_{1}<4\right) ^{\left[ K/\mu \right] }.
\end{equation*}
Thus $\Pr \left( \tau _{0,4}>K\right) \approx \left( 1-\Pr \left(
H_{1}=4\right) \right) ^{\left[ K/\mu \right] }$. Observing that $\mu $\ is
of order $1/\alpha ^{\prime }$, we get that the mean value of $\tau _{0,4}$\
is of order $1/\left( \alpha \alpha ^{\prime }\right) =e^{\left[ U\left(
2\right) -U\left( 0\right) \right] /\varepsilon }$\ with 
\begin{equation}
\Pr \left( \alpha \alpha ^{\prime }\tau _{0,4}>t\right) \rightarrow e^{-t/2}%
\text{ as }\varepsilon \text{ gets small}.  \label{exp}
\end{equation}
Thus the expected mean time to move from $\left\{ 0\right\} $\ to $\left\{
4\right\} $\ is the exponential of the global energy barrier $U\left(
2\right) -U\left( 0\right) $\ normalized by $\varepsilon $\ and the time $%
\tau _{0,4}$\ normalized by its mean converges in distribution to an
exponential distribution with mean $2$.

We can check that in the latter model $\pi _{0}+\pi _{1}=\pi _{3}+\pi _{4}$\
showing that the two stable state basins share the same weight.\newline

Suppose the states $\left\{ 0\right\} $\ $\left\{ 2\right\} $\ $\left\{
4\right\} $\ have a width, say $L_{0},$\ $L_{2}$\ and $L_{4},$\ where the
Markov chain undergoes a symmetric random walk before possibly attempting to
overcome the energy barrier. In this case, the mean values are expected to
behave like 
\begin{eqnarray*}
\left\langle \tau _{0,4}\right\rangle &\simeq &L_{0}L_{2}e^{\left( U\left(
2\right) -U\left( 0\right) \right) /\varepsilon } \\
\left\langle \tau _{4,0}\right\rangle &\simeq &L_{4}L_{2}e^{\left( U\left(
2\right) -U\left( 4\right) \right) /\varepsilon }
\end{eqnarray*}
including a factor involving the characteristic plateaux lengths of the
steady states. The walker has to overcome its energy barrier but also spends
some time in the flat regions $\left\{ 0,2\right\} $\ for the move $\left\{
0\right\} \rightarrow \left\{ 4\right\} $\ and $\left\{ 2,4\right\} $\ for
the move $\left\{ 4\right\} \rightarrow \left\{ 0\right\} .$\ The condition $%
L_{0}\neq L_{4}$\ introduces some skewness in the equilibrium weights of the
two stable state basins. These considerations are the discrete space-time
versions of the result known for a Langevin-type stochastic differential
equation evolving in a quartic double-well potential $U$\ with additive
white noise with small local variance $\varepsilon .$\ In this context, \cite
{VK}, if $a$\ and $b$\ are the stable states corresponding to a global
minimum of $U$\ and if $c$\ is the in-between unstable state 
\begin{eqnarray*}
\left\langle \tau _{a,b}\right\rangle &\simeq &\frac{1}{\sqrt{U^{\prime
\prime }\left( a\right) \left| U^{\prime \prime }\left( c\right) \right| }}%
e^{\left( U\left( c\right) -U\left( a\right) \right) /\varepsilon } \\
\left\langle \tau _{b,a}\right\rangle &\simeq &\frac{1}{\sqrt{U^{\prime
\prime }\left( b\right) \left| U^{\prime \prime }\left( c\right) \right| }}%
e^{\left( U\left( c\right) -U\left( b\right) \right) /\varepsilon }.
\end{eqnarray*}
\newline

Coming back to the previous symmetric case where $\left\{ 0\right\} $\ $%
\left\{ 2\right\} $\ $\left\{ 4\right\} $\ are `simple' states, we finally
address the following problem: what is the time $\widetilde{\tau }_{0,4}$\
needed to first hit state $\left\{ 4\right\} $\ starting from $\left\{
0\right\} $\ given the walker does not return to $\left\{ 0\right\} $\
again. Note that 
\begin{equation*}
\widetilde{\tau }_{0,4}=\tau _{0,4}-\sup \left( k<\tau _{0,4}:X_{k}=0\mid
X_{0}=0\right) .
\end{equation*}
We have $\widetilde{\tau }_{0,4}=1+\tau _{1,4}$\ where $\tau _{1,4}$\ is the
time needed to first hit state $\left\{ 4\right\} $\ starting from $\left\{
1\right\} $\ of the ergodic chain governed by the transition matrix on $%
\left\{ 1,..,4\right\} ^{2}$: $Q_{1,1}=0$, $Q_{1,2}=1$; $Q_{2,1}=1/2$, $%
Q_{2,3}=1/2$; $Q_{3,2}=\alpha $, $Q_{3,4}=1-\alpha $; $Q_{4,3}=\alpha
^{\prime }$, $Q_{4,4}=1-\alpha ^{\prime }.$\ For this $Q-$chain, the state $%
\left\{ 1\right\} $\ is now purely reflecting. Using the scale function $%
\varphi $\ of this new chain, $\Pr \left( H=4\right) =1/\varphi \left(
4\right) =\frac{1-\alpha }{2+\alpha }\rightarrow 1/2$\ ($\varepsilon
\rightarrow 0$). Similarly, the mean return time $\mu $\ to state $\left\{
1\right\} $\ tends to a finite value when $\varepsilon \rightarrow 0$\ so
that the mean value of $\widetilde{\tau }_{0,4}$\ tends itself to a finite
value when $\varepsilon \rightarrow 0$. Given there is no possible return to
state $\left\{ 0\right\} $, the mean time to first hit state $\left\{
4\right\} $\ turns out to be very short compared to $\tau _{0,4}$\ itself.

\end{document}